# Open Access and Discovery Tools:

# How do Primo Libraries Manage Green Open Access Collections?

François Renaville

*University of Liège Library*

The Open Access (OA) movement gains more and more momentum with an increasing number of institutions and funders adopting OA mandates for publicly funded research. Consequently, an increasing amount of research output becomes freely available, either from institutional, multi-institutional or thematic repositories or from traditional or newly established journals.

As of the end of April 2015, there are about 2,850 academic OA repositories (Green OA) of all kinds listed on OpenDOAR (http://www.opendoar.org). Scholarly OA repositories contain lots of treasures including rare or otherwise unpublished materials and articles that scholars self-archive, often as part of their institution's mandate (Harnad 2004). But it can be hard to discover this material unless users know exactly where to look.

Since the very beginning, libraries have played a major role in supporting the OA movement. Next to all services they can provide to support the deposit of research output in the repositories, they can make OA materials widely discoverable by their patrons through general search engines (Google, Bing, etc.), specialized search engines (like Google Scholar), and library discovery tools, thus expanding their collection to include materials that they would not necessarily pay for.

In this paper, we focus on two aspects regarding Green OA and the Primo discovery tool.

In early 2013, Ex Libris Group started to add institutional repositories into Primo Central Index (PCI), their mega-aggregation of hundreds of millions of scholarly e-resources (journal articles, e-books, reviews, dissertations, legal documents, reports, etc.) (http://www.exlibrisgroup.com/category/PrimoCentral). PCI is an additional service available to the Primo discovery tool customers. After two years, it may be interesting to take stock of the situation of PCI regarding OA institutional repositories.

On the basis of a survey carried out among members of the Primo community, we will see how libraries using Primo discovery tool integrate Green OA contents in their front-end. Two major ways are possible for them. Firstly, they can directly harvest, index, and manage any repository in their Primo instance and display those free materials next to the more traditional library collections. Secondly, if they are PCI subscribers, they can quickly and easily activate any, if not all, of the 74 OA repositories contained in PCI, making thus the contents of those directly discoverable to their end users.[1] This paper shows what way is preferred by libraries, if they harvest or not their own repository (even if it is included in PCI) and suggests efforts that Ex Libris could take to improve the visibility and discoverability of OA materials included in the "Institutional Repositories" section of PCI.

## Survey Results

The survey contained multiple-choice and open questions related to OA local sources and the usage and perception of the "Institutional Repositories" section of PCI. It was posted to the listserv PRIMO-DISCUSS-L in March 2015 and was open from March 16 until April 10, 2015. PRIMO-DISCUSS-L has about 1,500 subscribers, who are typically local Primo administrators or managers. As it was an institutional survey, only one response per institution was expected.

The survey received 34 responses from 15 countries: Australia (2), Austria (1), Belgium (1), Brazil (1), Canada (2), Denmark (1), France (5), Iceland (1), Netherlands (2), New Zealand (1), Norway (1), Sweden (1), Switzerland (3), United Kingdom (5), and United States (7).

## Harvesting of Local OA Collections

Of the 34 respondents, 20 (59%) have said they harvest a local institutional repository (IR) in their Primo instance, 7 did not and 7 did not at the moment of the survey, but planned to do so.

Respondents were also asked if all the records they harvest (or plan to) have a least one OA file (Figure 1). Of the 27 concerned respondents, 6 said it is the case and 6 admitted they didn't know. For most of the respondents, not all local records they harvested have an OA file, to varying degrees. However, harvesting few OA contents do not necessarily mean harvesting few records: some large repositories, especially if they also act as an institutional bibliography, may have more OA content than smaller repositories with a higher percentage of OA content.

---

[1] As of April 27, 2015.

An institutional bibliography aims to offer a web based instrument to capture, proceed, use and disseminate bibliographic information of the output and ongoing research of a university or research institution. One of its goals is to increase the external and internal visibility of the research done. Of the 27 respondents who harvest or plan to harvest an Institutional Repository (IR), 15 have admitted that their repository also acts as an institutional bibliography, at least partially for some departments.

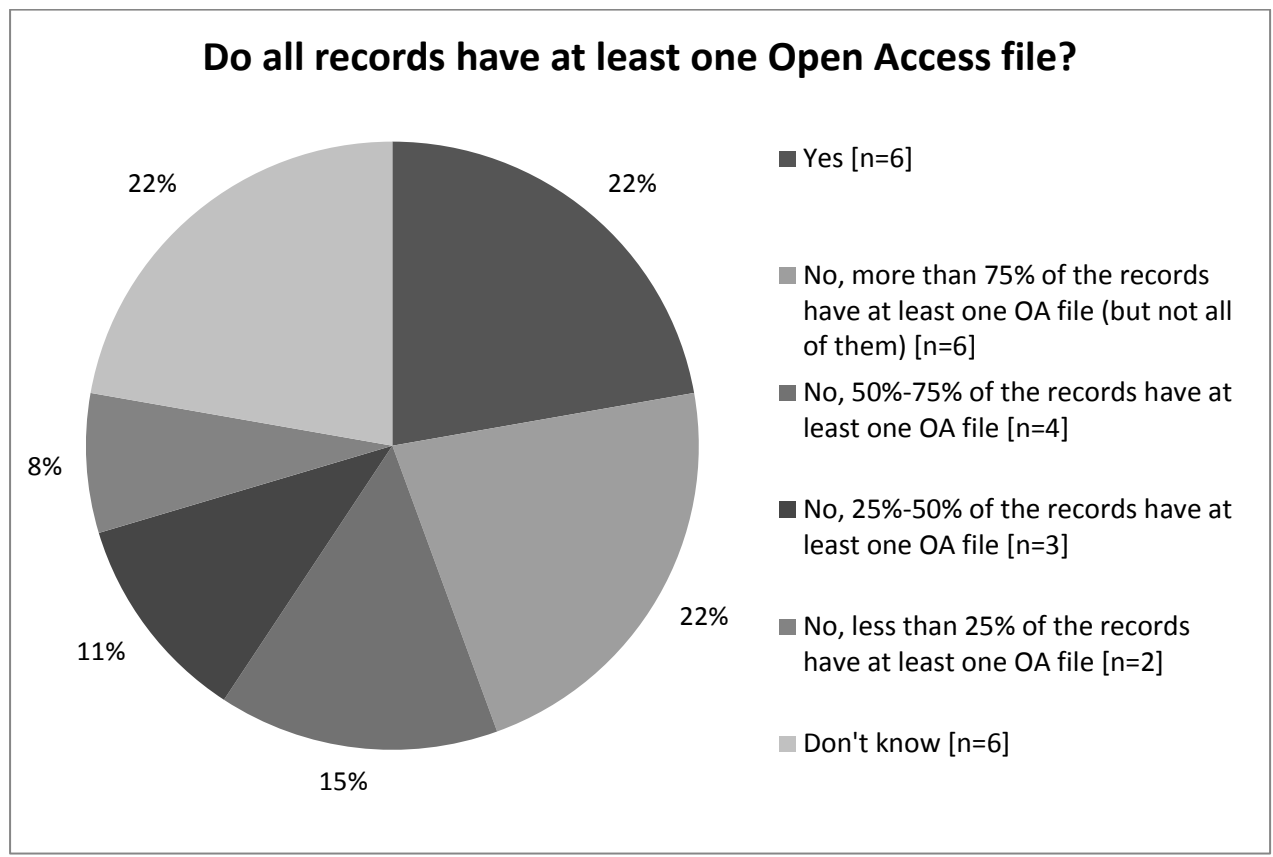

*Figure 1: Percentage of records with at least one OA file*

Of the 34 respondents, 17 said they also harvest or plan to harvest an additional OA repository in their Primo for their end-users. Examples given are various: it may be other selected traditional OA repositories (2 respondents), or OA repositories with contents of research or regional interest, but not necessary with scholarly research output (e.g., "University's repository of learning objects"; "archival photos, videos of researchers talking about their work, old university publications […] including Creative Commons licenses for all material"; "digitized manuscripts and rare books, pictures, videos, sound recordings… which belong to our institution or to partner institutions within the country and are freely available to all").

Survey participants were also asked if they have defined or intend to define any specific configuration in the Primo Back Office administration for their local OA scholarly materials in order to prominently display their OA institutional research output within the results of the discovery tool. Five methods for achieving this goal were suggested to the respondents:

1. Boosting: The Primo discovery tool allows the library to negatively boost records that are not from the institution. In other words, PCI records or records from a partner institution (e.g. in case of consortium) may be placed below records of the institution in the Brief Results page. However, this option does not make any distinction between bibliographic records coming from the ILS and records coming from any harvested repository, both sources being local data.

2. Indexes: Particular indexes for OA local contents can be created in the Simple or Advanced Search interface.

3. Display: Any display configuration like label, logo, specific information in the Details tab.[2]

4. Links: Primo allows the library to display additional links in the full bibliographic record description. In case of OA content, such links could, for example, lead to the original record, to the full text, to copyright information, etc.

5. Facets: Facets are links in the Brief Results page that allow users to filter their search results by a specific category, such as creator, language, topic or, in this particular case, for institutional OA materials.

Respondents could also enter text into an "other" option.

Figure 2 shows the frequency of these five responses from the 27 respondents who harvest or plan to harvest a scholarly OA repository: Seven answered that they use boosting options so that local data may be displayed before PCI records. Two will use specific indexes. Five use particular display rules, for example by displaying a Creative Commons license where applicable on the Details tab or by adding a line like "Open Research Online - a research publication from the University" to the Brief Results page. Six use links to promote their OA contents ("The associated OA files appear in the links section of detailed view pages"; "Direct link to the reference via View It + additional link to the reference in the Details tab").[3] Finally, 11 respondents answered that they use facets; 2 mentioned that they promote their local OA research output with a top-level institutional facet.[4] Moreover, one respondent has explained they have created a specific search scope containing part of the contents of their OA repository (theses and dissertations), one more intends to create an additional scope for their harvested OA research output.

---

[2] In Primo, the Details tab displays the item's full record and additional links.

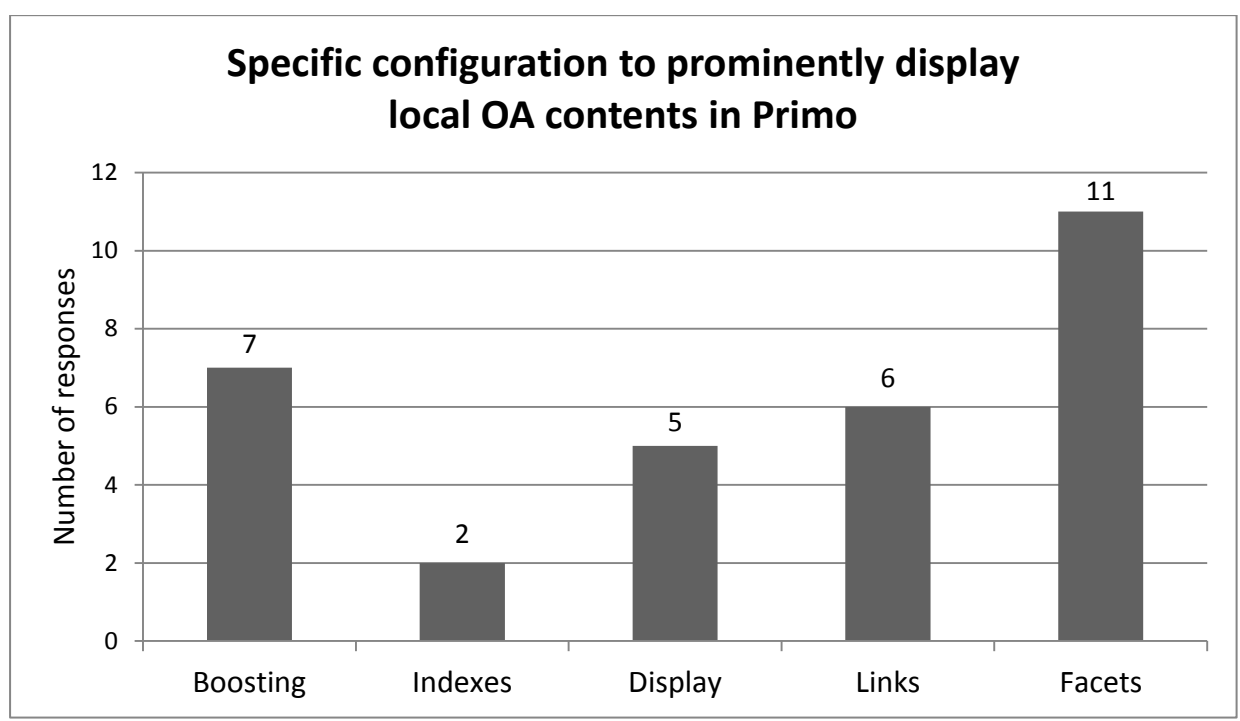

*Figure 2*: *Specific configuration to prominently display local OA contents in Primo*

Finally, one respondent intends to combine four of these promoting options (links, facets, display, indexes), two use three of them (boosting, facets and links), and six use or will use a combination of two means.

## Primo Central Index

Primo Central Index (PCI) is the name for Ex Libris' mega-index of "hundreds of millions of scholarly e-resources of global and regional importance. These include journal articles, e-books, reviews, legal documents and more that are harvested from primary and secondary publishers and aggregators, and from open-access repositories."[5] Those records are mainly provided by publishers (for example, Wiley, Springer, Elsevier, and Thomson/Reuters), who provide the metadata from their publication platforms (Wiley Online Library, SpringerLink, ScienceDirect, Scopus, Web of Science) to PCI, and by aggregators (e.g., ProQuest) and their databases. PCI allows direct access to the metadata but not full texts, which are only available to the customers and their end-users if the institution subscribes to that service. Therefore, PCI works in close relationship with link resolvers (like SFX) and makes it possible to display to end-users only records for which an access to the full text is available. In addition to publishers' and aggregators' content, Ex Libris has also included in PCI large collections and archives of free scholarly materials: *ArXiv.org*, *HAL (Hyper Article en Ligne)*, *OAPEN: Open Access Publishing in European Networks*, *SwePub*, *Norwegian Open Research Archives.*

---

[3] In Primo, the View Online Tab (or View It) displays items that are available online inside the tab.
[4] In Primo, top-level facets are static facets that display in the "Show only" section (first section above) of the Brief Results page. Unlike other kinds of facets, top-level facets always display even if there is only one matched record in the category.

In January 2013, Ex Libris "released a new service for institutional –and open access– repositories [in order to] simplify the process of allowing their content to be indexed in Primo Central" (Ex Libris Group 2013a). That registration service was open to all institutions, not only to Primo customers. The goal was to enable users at Primo institutions to discover more easily such OA materials. That IR registration service was part of a wider Ex Libris initiative to support OA. In addition to adding more OA material to their indexes, Ex Libris worked also on improving access to OA articles in subscription (hybrid) journals (Ex Libris Group 2013a).

Activating collections in PCI is a very easy task. A widget displays all the available collections, sorted by publisher, with brief descriptive information related to the update frequency and the day the resource was included in PCI. Some collections are restricted for search and delivery: they may only be activated by the customer if the institution has a valid subscription at the original provider (e.g., *Scopus* [Elsevier], *Web of Science* [Thomson/Reuters], *MLA Institutional Bibliography* [MLA], ProQuest databases [ProQuest], *L'Année Philologique* [Société internationale de bibliographie classique], *GeoRef* [American Geosciences Institute]).

---

[5] Ex Libris Group. Primo Central Index. http://www.exlibrisgroup.com/category/PrimoCentral

## Institutional repositories in PCI

Following the release of the registration service, the first IRs were integrated and available to PCI subscribers in March 2013. Two years later, at the end of March 2015, 74 IRs are integrated in PCI.

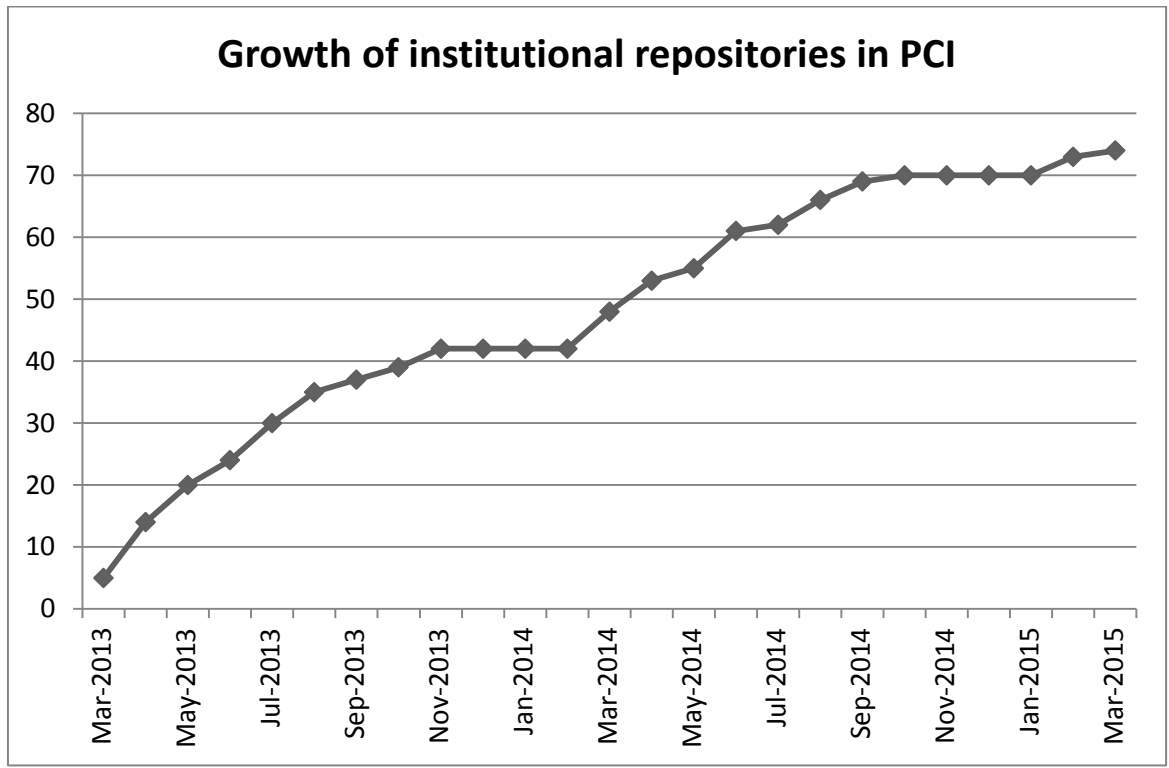

*Figure 3: Growth of institutional repositories in PCI (from March 2013 to March 2015)*

As Figure 3 shows, the number IRs has mainly increased from spring to autumn while it stagnates at the end and the beginning of the year. However, two years after the new registration service had been launched, the service link is surprisingly not available anymore and has disappeared from the Ex Libris website.[6]

The 74 IRs come from 19 countries: United Kingdom (17), United States (14), Spain (7), Germany (6), Australia (4), Canada (4), Belgium (3), Netherlands (3), Norway (3), Czech Republic (2), Peru (2), Switzerland (2), Austria (1), Brazil (1), Ireland (1), Italy (1), New Zealand (1), Sweden (1), and Taiwan (1).

---

[6] However, the service page can be found with the help of the Internet Archive's Wayback Machine (http://web.archive.org/web/20130509051917/http://dc02vg0047nr.hosted.exlibrisgroup.com:8080/IRWizard/wizard.html).

## Resource type and number of records

PCI aggregates various sources with various contents.[7] Therefore, all original resource types from the IR are mapped by Ex Libris with one of the 23 PCI resource types:

- Article
- Audio
- Book
- Book Chapter
- Conference Proceeding (includes proceedings volumes as well as individual papers)
- Database
- Dissertation (PhD and master thesis)
- Government Document (publications issued by government agencies, including patents, excludes court opinions, case briefs, etc.)
- Image
- Journal (includes journal issues)
- Legal Document (court opinions, case briefs, etc.)
- Map
- Newspaper Article
- Reference Entry (individual entries in dictionaries, encyclopedias)
- Research Dataset (raw data produced as the output of research)
- Review (book, product, film reviews)
- Score
- Statistical Data Set (tables, graphs containing statistics)
- Technical Report
- Text Resource (unclassifiable textual sources)
- Video
- Website
- Other (unclassifiable non-textual sources)

For library administrators in charge of Primo, successfully managing those resource types in relation with their local data is not always an easy task (Koster 2012).

---

[7] See details in Appendix.

Among those resource types, "Text Resource" is a particular one. Ex Libris relies on the metadata given to them by the provider. Some providers supply appropriate resource types like "article" or "book chapter" while some providers simply state that the resources are "text." Therefore, depending on the quality of the provided metadata, many books, articles, reports, reviews, etc., may be hidden behind a generic and unclear "Text Resource" which represents 29% of all materials coming from IRs (Figure 4).

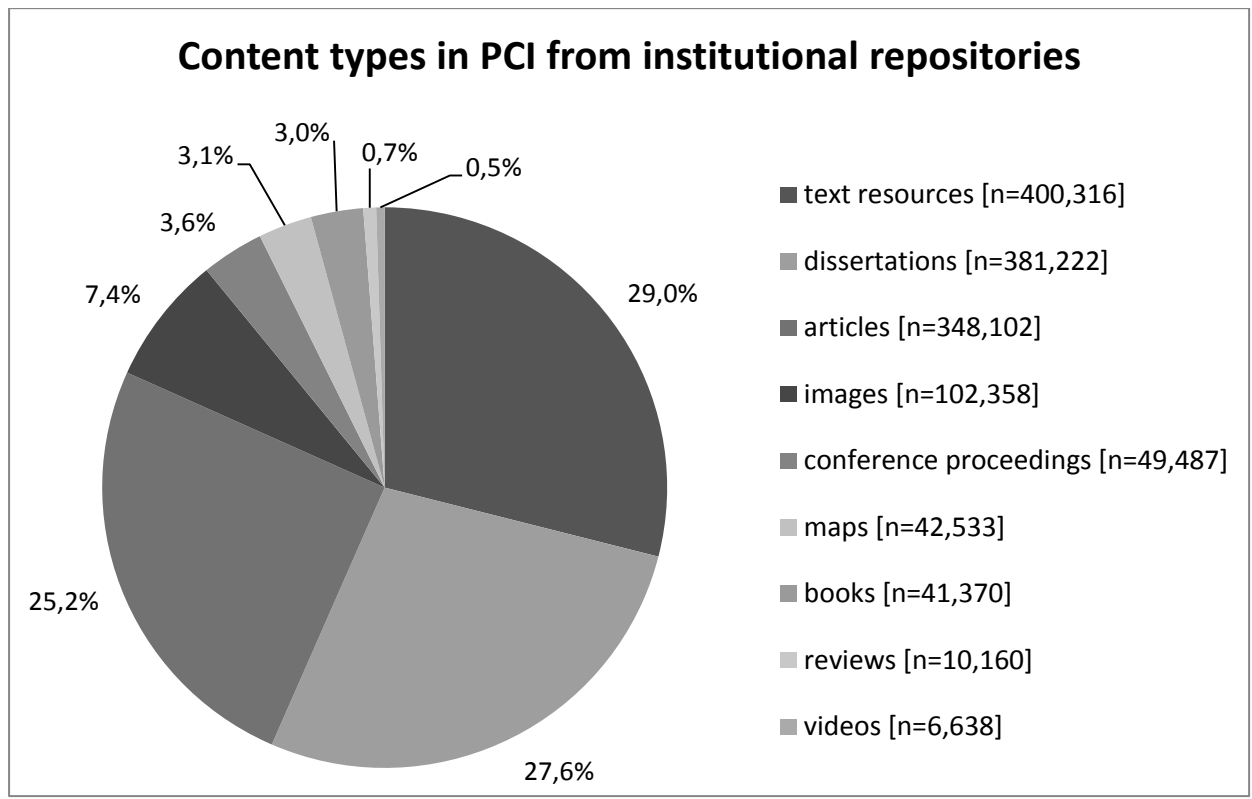

*Figure 4: Main resource types and number of records (March 31, 2015)*

After text resources, dissertations and articles are the most two frequent materials provided by IRs. In some cases, those contents represent the entire harvested content of repositories (Figures 5 and 6).

## Top 10 Institutional Repositories in PCI - Dissertations

| Institutional Repositories | dissertations | % of the harvested IR content |
|---|---|---|
| DiVA - Academic Archive Online (Uppsala University Library) | 138,761 | 76.1% |
| ETDs Repository (VŠKP - University of Economics, Prague) | 37,757 | 99.6% |
| BRAGE (BIBSYS) | 33,836 | 45.1% |
| National Chung Hsing University Institutional Repository (National Chung Hsing University, Taiwan) | 28,168 | 41.9% |
| ThinkTech (Texas Tech University) | 17,622 | 90.7% |
| Dokumentenserver der FU Berlin (Freie Universität Berlin) | 10,587 | 100.0% |
| Digital Dissertations (Universitätsbibliothek der LMU München) | 10,497 | 100.0% |
| Digital Library of the University of Pardubice | 8,399 | 80.7% |
| Diposit Digital de Documents de la UAB (Universitat Autonoma de Barcelona) | 6,771 | 11.5% |
| RiuNet (Universitat Politècnica de València) | 6,500 | 24.0% |

**Figure 5**: *Top 10 Institutional Repositories in PCI - Dissertations (March 31, 2015)*

## Top 10 Institutional Repositories in PCI - Articles

| Institutional Repositories | articles | % of the harvested IR content |
|---|---|---|
| Repository Utrecht University | 38,717 | 57.9% |
| Infoscience (École Polytechnique Fédérale de Lausanne (EPFL)) | 30,762 | 100.0% |
| Diposit Digital de Documents de la UAB (Universitat Autonoma de Barcelona) | 21,189 | 36.1% |
| ORBi (Open Repository and Bibliography) (University of Liège) | 19,665 | 49.2% |
| ZORA (University of Zurich) | 19,152 | 70.6% |
| National Chung Hsing University Institutional Repository (National Chung Hsing University, Taiwan) | 17,499 | 26.0% |
| VU-DARE (VU University Amsterdam) | 16,271 | 66.8% |
| DiVA - Academic Archive Online (Uppsala University Library) | 15,431 | 8.5% |
| Digital Access to Scholarship at Harvard (DASH) (Harvard University, Office for Scholarly Communication) | 15,076 | 80.9% |

| Lirias (KU Leuven Association) | 12,736 | 49.0% |
| --- | --- | --- |

*Figure 6: Top 10 Institutional Repositories in PCI - Articles (March 31, 2015)*

As of March 31, 2015, PCI contained more than 1,600,000 records coming from IRs. According to Ex Libris policy, all those included records should provide access to an OA full text. The rule is that "[o]nly content which is 100% Open Access should be included in the Institutional Repositories collections" (Ex Libris Group 2014). In the case a repository is not fully OA (e.g. in case of an institutional bibliography) it is possible to ingest only the records which contain specific values in an OA tag so that "[o]nly records which include links to unrestricted Full Text or to a publishers landing page where unrestricted Full Text is available" (Ex Libris Group 2014) are indexed in PCI.

However, experience and remarks from survey participants (see below) have revealed this was far from the case. It should also be emphasized that at the very beginning candidate repositories were not required to be entirely OA, or that metadata must not necessarily include clear consistent OA indicators. The very first prerequisites for submitting an IR for inclusion in PCI (Ex Libris Group 2013b) were indeed:

- *"Content should be of scholarly interest*
- *Content should contain sufficient metadata using standard formats such as Dublin Core or MODS*
- *Initially we will be focusing on OAI-PMH as the harvesting method".*

These soft and very basic requirements from the beginning may partially explain why some content provided by repositories may be without any OA file.

## Usage and perception of PCI

Institutions that have an institutional repository whose content is harvested in PCI *and* also subscribe to the PCI service may choose to directly harvest and index in their Primo instance or to activate their own repository in the Primo admin interface.[8] When activating one's own repository via PCI, the institution benefits from automatic weekly updates. New OA content coming from the repository is added in PCI and made available via the institution's Primo. On the other side, directly harvesting its own repository allows more flexibility, but requires more investment from the library staff. For example, specific configuration in order to prominently display local OA contents in Primo is possible (see above). The institution has also the possibility to include repository items with restricted access (campus only) and to make them searchable via Primo for end users, and more frequent updates can be scheduled. On the other hand, that flexibility requires from local Primo administrators to carry on and monitor the whole harvesting and indexing process.

---

[8] Theoretically, they can also do both: activating their IR in PCI for OA contents and directly harvesting records with no OA full text, but this is additional work and may display duplicates in the front end.

All survey participants (34) belong to an institution that subscribes to PCI. Of these, 18 include their own institutional OA materials in PCI; for 14 of them, the local IR is directly harvested by PCI. In the 4 other cases, even if the IRs are not directly harvested in PCI, they are actually *indirectly* harvested since their OA materials are harvested by another resource such as a national scholarly repository, which is then harvested by PCI. For 10 respondents, the OA repository is not yet included in PCI; 2 of them already requested Ex Libris to add it. The last 6 respondents have no institutional repository that is harvested by PCI at the time of the survey.

The survey aimed to determine whether libraries whose IR is harvested by PCI prefer to activate it in their PCI administrative interface and by doing so effortlessly get their own OA materials available in Primo or if they prefer instead to harvest their IR. This second option requires some additional work for Primo administrators, but it allows a more flexibility and more frequent harvesting and indexing of new and updated records archived in the repository and the possibility of putting forward local OA materials (boosting, specific indexes, particular facets, scopes).

Figure 7 shows that of the 14 respondents whose IR is harvested by PCI, 7 have not activated it in PCI, while 7 have. Surprisingly, of the 7 who have activated their IR in PCI, 5 also directly harvest their local IR in their own Primo. Of the 7 who have not activated their own IR PCI, 6 directly harvest it and one does not.

**Repository in PCI: activated or not? Directly harvested or not?**

|                     | Activated in PCI | Not activated in PCI |
|---------------------|:----------------:|:--------------------:|
| Directly harvested  | 5                | 6                    |
| Not harvested       | 2                | 1                    |
| Total               | 7                | 7                    |

*Figure 7: Repository in PCI: activated or not? Directly harvested or not?*

One respondent also explained they activated their IR in PCI, but only for a brief testing period. They considered that information presented in Primo via PCI was not good enough and regretted that "all resources were considered available" while their "default configuration of Primo is set to display available resources only."[9] Another respondent (a PCI subscriber whose IR is not included in PCI) noticed a similar defect, arguing that "PCI cannot differentiate between no [OA] full text and [OA] full text records." According to Ex Libris' basic requirements (see above), this should not occur.

Of the 34 PCI subscribers, 19 (56%) admitted that they decided not to activate some IRs included in PCI or to deactivate others. The reasons they gave are closely related to the potentially negative effect that those unwanted collections could have on their users. Seven respondents (21%) explained that for some collections there is not always an OA full text available and that full text –if there is any– is only available after logging in on the original platform (restricted access):

- "They were causing errors in Primo search results, such as showing up as available, until you tried to access them and then you get a "no full text available" message or it turns out to be just a record with no full text."

- "When it appears that a record is only accessible when logged in, we deactivate the collection. This happens quite regularly."

The second most frequent reason, given by 5 respondents (15%), is that some IRs might be deactivated if their content is considered to be not relevant or out of scope for the end-users:

- "To prevent from noise, lack of interest for some contents"

---

[9] However, this is configurable in the Primo admin interface with a slight adaptation of the Real-Time Availability (RTA) rules. By default, the RTA feature allows Primo sites to get real-time availability statuses for physical items directly from the source system, but local Primo administrators can also use them to display a particular availability text for repository records with no OA full text.

- "Some repositories which do not content real scholarly production –like images related to the history of a particular institution– or whose content is for a very local usage, and of very few interest for our users."

Two respondents mentioned both reasons.

Survey participants were also asked if they applied a specific selection policy before activating new resources added to the PCI "Institutional Repositories" section (Figure 8). Four main policies (direct activation, analysis based on the content, on the availability of open access content, and on the metadata quality) and their combinations were proposed.

| Selection policy | Number | Percentage |
|---|---|---|
| (a) We usually directly activate any new repository in the Production environment (or in both Staging and Production environments) (no checking). | 10 | 29% |
| (b) We first activate new repositories in the Staging environment, and activate in the Production after we have controlled that its content is appropriate for us (discipline oriented). | 1 | 3% |
| (c) We first activate new repositories in the Staging environment, and activate in the Production after we have controlled that its content is (mainly) in Open Access. | 0 | 0% |
| (d) We first activate new repositories in the Staging environment, and activate in the Production after we have controlled that metadata are good enough. | 0 | 0% |
| Combination: (b) + (c) | 3 | 9% |
| Combination: (b) + (d) | 1 | 3% |
| Combination: (c) + (d) | 2 | 6% |
| Combination: (b) + (c) + (d) | 2 | 6% |
| (e) We rarely or never activate resources from the "Institutional Repositories" section. | 7 | 21% |
| (f) Other workflow | 8 | 24% |
| Total | 34 | 100% |

*Figure 8: Specific selection policy before activating new IRs in PCI*

While 21% of respondents complained that there is not systematically OA content in records from IRs –and this in contradiction with Ex Libris' policy– that same percentage is reflected in the selection workflow that libraries have set up. Surprisingly, more than 1 in 5 respondents concede they rarely or never activate resources from the "Institutional Repositories" section.

Two respondents explain they had not considered such OA collections carefully yet and that it is a project that is underway, in conjunction with an analysis and revision of their entire collection of subscriptions (periodicals and databases). Another argues that library policy prevents activating those collections, while a second explains they "have an ongoing debate on displaying OA resources in [their] discovery system." Finally, one last respondent explains it is a content and access issue ("Content often not relevant to our specialist institution. Percentage of OA material would also need testing - don't want something coming up as full text access and then users clicking through and finding it's not full text after all").

Among those who have another workflow, one respondent explains they have set up an internal decision making process, in association with other librarians, since in their opinion the Primo administrator should not be alone in activation decision. For another, they directly activate in their production environment any new repository that has appropriate content, without checking on the staging server before. Finally, one respondent specifies that since they are part of a consortium, their PCI is centrally managed by the consortium office.

Among the respondents who regularly activate OA resources in PCI's "Institutional Repositories" section, 5 do not see any repository that should be removed from PCI. On the other hand, 5 respondents are not against that possibility. The reason that is put forward by 4 of them is related to those IRs whose harvested results do not always contain an OA file:

> *"Some resources have too many false-positives (online access without any full text). It is frustrating to the users and gives us a lot of error reports to handle manually."*
> *"That IR section should only contain repositories that provide 100% OA content to Ex Libris. Unfortunately, it is not always the case and it is terribly frustrating for the end-users."*
> *"If there is no link to full test from the view it tab, it frustrates our users, or if there is no full text available at all. We might consider de-activating if we get recurrent complaints."*
> *"All of those which require an institutional login."*

Survey participants were also asked to give their opinion about the necessary efforts that Ex Libris should make to improve the "Institutional Repositories" section of PCI. Seventeen respondents made different suggestions which can be spread into 6 categories. Many enhancement requests concern repository records where OA availability is incorrectly indicated (false positives) and that should normally not be made available through PCI:

1. Providing *only OA contents* in the "Institutional Repositories" section or at least making more checks for the availability of full text (systematic monitoring), so that libraries can be sure they will activate only the repositories which have at least one OA file attached (pointed out 8 times).

2. Providing *more detailed information* about each OA repository for local PCI administrators (number of records, percentage of eventual non OA entries [sic], types of documents, disciplines…) (pointed out 3 times)

3. Adding *more OA contents* (by improving the awareness within the international academic community about the existence of harvesting activity repositories by Ex Libris, by adding more available free resources such as national library collections, and special governmental collections) (pointed out 3 times).

4. Requiring OA content providers to provide *good metadata quality* and to adhere to OA indicator standards (for example as defined by a standards body like NISO) (pointed out twice) (NISO 2014).

5. Establishing a *better collaboration* between Ex Libris and the host of the metadata to ensure that the content can quickly and correctly be integrated into PCI (for example the final result could be approved by the metadata provider before the new OA repository is available in PCI for other customers) (pointed out twice).

6. Allowing richer and more extended *record formats* for harvesting purposes (DIDL, MODS…) (pointed out twice)

When asked about the general efforts that could be deployed to bring together OA content so that it is not so laborious to add and make it easily and quickly visible for end-users, 5 respondents suggested improving the way OA collections are displayed in the front end. Respondents suggested the use of consistent facets for OA content or specific OA flags (for example in the Brief Results) like the <free_to_read> XML element "to indicate that content can be read or viewed at its current location by any user without payment or authentication" (NISO 2015). Improvement can also be made in the back-end interface by getting more information about the various collections which are regularly added in PCI (e.g. number of records, types of documents included, thematic coverage, etc.) so that it becomes easier for local Primo administrators to determine which collections should be activated or not. Those expectations are confirmed by a recent survey (Bulock 2015):

> *"Another common theme […] involved frustration with inaccurate information in OA management systems affecting patron discovery and staff workflows.*

> *Hosburgh noted that databases and discovery services often include facets designed to limit result sets to OA items. Librarians praised knowledgebases for helping them provide more OA resources, and appreciated link resolvers that clearly labeled OA items as "FREE" on the result screen."*

One respondent suggested the possibility of choosing a rank of boosting (of results) for each resource, and not only for local contents vs PCI records. One respondent also pointed out that smart grouping of records is "very important, and even more for green OA where different versions of the same document can be offered from various sources (publisher platform, one or more institutional repository/ies or subject repository...) with varying metadata quality."[10] Finally, two respondents recommended also that Ex Libris harvest first and foremost national or global repositories instead of smaller repositories with less content.

## Conclusions

The survey shows that most of the respondents (59%) harvest a local institutional OA repository in their Primo instance. Several have defined specific configurations in their Primo in order to put forward their institutional OA research output within the results of the discovery tool; facets and boosting local data being the two most popular ways. All survey participants belong to an institution that subscribes to PCI, and for 14 of them their local IR is directly harvested by PCI. Some respondents have admitted that in addition to directly harvesting their repository they have also activated their own IR in PCI.

All survey participants are also PCI subscribers. 56% of them admitted that there are in PCI some IRs they decided not to activate, or to deactivate. The two main reasons are related to potential negative effects that those collections could have on their end-users: materials of some IRs are considered to be not relevant for the end-users and, for some IRs collections, there is not always an OA full text available, contrary to what Ex Libris requires from repository content providers. One in 5 respondents confessed they always check the availability of announced OA contents before activating them in the production environment.

---

[10] In PCI, grouping is based on an existing FRBR workflow that is used to prevent duplicate records in the results list. In the Primo front end, the system displays the preferred record in the brief results list. Preference is always given to the original publisher's record over an aggregator record, a pre-print or a record coming from an institutional repository. Users can easily access to additional versions by clicking the "View all versions" link.

In terms of number of records, the "Institutional Repositories" section of PCI *theoretically* allows subscribers to display more than 1,600,000 OA records, of which about 350,000 are scholarly articles and 380,000 are dissertations. Even if 21% of the respondents concede they rarely or never activate resources from the "Institutional Repositories" section in PCI, many participants stress that improvement is necessary for those materials, notably by providing *only* OA contents in the "Institutional Repositories" section, by providing more detailed information about each OA repository, and by adding more OA repositories. Improving the way OA collections are displayed in the front-end is also encouraged. By improving *quality* and *quantity* of the "Institutional Repositories" section in PCI, Ex Libris can certainly hope to gain gratitude from their customers.

*First version: April 2015*
*Revised: June 2015*

Bibliography:

- Bulock, C., Hosburgh, N., Sanjeet, M. 2015. OA in the Library Collection: The Challenges of Identifying and Maintaining Open Access Resources, *The Serials Librarian: From the Printed Page to the Digital Age* 68(1-4), 79-86. doi: 10.1080/0361526X.2015.1023690

- Ex Libris Group. 2013a, January 31. Institutional Repositories in Primo Central: Registration is now open. [Blog post]. http://initiatives.exlibrisgroup.com/2013/01/institutional-repositories-in-primo.html

- Ex Libris Group. 2013b. Institutional Repositories Registration. http://web.archive.org/web/20130509051917/http://dc02vg0047nr.hosted.exlibrisgroup.com:8080/IRWizard/wizard.html [Capture from May 9th, 2013, by Wayback Machine]

- Ex Libris Group. 2014. *Institutional Repository Submission Requirements*.

- Harnad, S., Brody, T., Vallières, F., Carr, L., Hitchcock, S., Gingras, Y., Oppenheim, C., Stamerjohanns, H., Hilf, E. 2004. The Access/Impact Problem and the Green and Gold Roads to Open Access. *Serials Review* 30 (4): 310–314. doi: 10.1016/j.serrev.2004.09.013

- Koster, L. 2012. Wizards and Tables or how the Primo back office could work better. Paper presented at *IGeLu 2012 Zurich*. http://igelu.org/wp-content/uploads/2012/09/Wizards-and-Tables_Koster.ppsx

- NISO. 2014. *Open Discovery Initiative: Promoting Transparency in Discovery*. http://www.niso.org/apps/group_public/download.php/14821/rp-19-2014_ODI.pdf

- NISO. 2015. *Access License and Indicators. A Recommended Practice of the National Information Standards Organization*. http://www.niso.org/apps/group_public/download.php/14226/rp-22-2015_ALI.pdf

**Appendix**

The number of records per resource type was retrieved from a Primo instance where all IRs and been activated at least two weeks earlier in Primo Central Index. It was not possible to get any results for "Digital Repository @ Iowa State University (Iowa State University)" and "eScholarship (University of California, California Digital Library)." Therefore, the following table contains results for 72 repositories. Only the results for the 9 most common resource types (article, book, conference proceeding, dissertation, image, map, review, text resource, and video) are presented, all other resource types (database, journal, legal document, newspaper article, reference entry, research dataset score, technical report, statistical data set, website, and other) have been grouped into one general category (all others).

Searches were made in two steps on March 31, 2015. First, for each repository, meaningful keywords were used to retrieve a maximum number of potential records:

- "Utrecht" to search for "Repository Utrecht University"
- "Leeds University" to search for "Leeds Met Open Search (Leeds Metropolitan University)"
- "London School of Economics" to search for "LSE Research Online (London School of Economics and Political Science)" and "LSE Theses Online (London School of Economics and Political Science)".

Secondly, collection facets were used to retain only records coming from the IRs. It was not possible to use facet filtering for four repositories:

- Manchester eScholar (Manchester University)
- Opus: Online Publications Store (University of Bath)
- ETDs Repository (VŠKP - University of Economics, Prague)
- Dokumentenserver der FU Berlin (Freie Universitat Berlin)

However, with the used search terms ("Manchester eScholar", "Online Publications Store" AND "University of Bath", "VŠKP", and "Dokumentenserver der FU Berlin") and their results, there was no room for uncertainty.

In some cases, the total number of records is lower than the sum of all resource types. The reason is that when multiple records from different sources (IRs or not), with a different resource type, are FRBRized, facets take the different values into account (for instance, if an item that is an article is grouped with an item which is a review, both of them will have both facet values).

| Institutional Repositories | Total number of records | Number of records per resource type | | | | | | | | | |
|---|---|---|---|---|---|---|---|---|---|---|---|
| | | article | book | conf. proceed. | dissert. | image | map | review | text resource | video | all others |
| ARAN (National University of Ireland Galway) | 3,331 | 1,262 | 155 | 506 | 720 | 1 | 0 | 36 | 481 | 0 | 894 |
| Bergen Open Research Archive (University of Bergen) | 7,578 | 1,892 | 52 | 101 | 5,303 | 1 | 0 | 11 | 401 | 4 | 0 |
| BRAGE (BIBSYS) | 75,026 | 6,842 | 1,075 | 492 | 33,836 | 14 | 2 | 7 | 22,788 | 63 | 4 |
| CERES (Cranfield Collection of E-Research) (Cranfield University) | 445 | 400 | 0 | 28 | 0 | 0 | 0 | 10 | 7 | 0 | 1,210 |
| Constellation (Université du Québec à Chicoutimi) | 2,436 | 137 | 106 | 21 | 2,015 | 0 | 0 | 67 | 88 | 0 | 848 |
| Die digitale Landesbibliothek Oberösterreich (Upper Austrian Federal State Library) | 3,515 | 3,515 | 0 | 0 | 0 | 0 | 0 | 0 | 0 | 0 | 0 |
| Digital Access to Scholarship at Harvard (DASH) (Harvard University, Office for Scholarly Communication) | 18,629 | 15,076 | 525 | 780 | 1,715 | 0 | 0 | 186 | 2,440 | 0 | 255 |
| Digital Commons @ SPU (Seattle Pacific University) | 3,017 | 0 | 0 | | 0 | 124 | 0 | 0 | 2,222 | 671 | 2,155 |
| Digital Dissertations (Universitätsbibliothek der LMU Muenchen) | 10,497 | 0 | 0 | 0 | 10,497 | 0 | 0 | 0 | 0 | 0 | 74 |
| Digital Library of the University of Pardubice | 10,412 | 1,451 | 21 | 522 | 8,399 | 0 | 0 | 0 | 23 | 0 | 5 |
| DigitalCommons@McMaster (McMaster University Library) | 13,310 | 12,231 | | | | | | | | | 44 |
| Digitale Sammlungen (Universitätsbibliothek Paderborn) | 3,191 | 639 | 1,632 | 0 | 867 | 0 | 0 | 0 | 24 | 0 | 0 |

| | | | | | | | | | | |
|---|---|---|---|---|---|---|---|---|---|---|
| Diposit Digital de Documents de la UAB (Universitat Autonoma de Barcelona) | 58,675 | 21,189 | 571 | 397 | 6,771 | 0 | 0 | 0 | 16,756 | 0 | 0 |
| Diposit Digital de la Universitat de Barcelona | 16,462 | 7,219 | 272 | 40 | 5,444 | 189 | 0 | 16 | 2,955 | 48 | 0 |
| DiscoverArchive (Vanderbilt University) | 6,241 | 815 | 137 | 0 | 323 | 138 | 5 | 6 | 1,063 | 2,287 | 0 |
| DiVA - Academic Archive Online (Uppsala University Library) | 182,337 | 15,431 | 5,624 | 6,984 | 138,761 | 0 | 0 | 1,508 | 12,864 | 0 | 892 |
| Dokumentenserver der FU Berlin (Freie Universität Berlin) | 10,587 | 0 | 0 | 0 | 10,587 | 0 | 0 | 0 | 0 | 0 | 345 |
| DSpace@Cambridge (University of Cambridge) | 188,334 | 229 | 7 | 2 | 17 | 5,223 | 2 | 0 | 6,823 | 0 | 9,932 |
| DUGiDocs (Universitat de Girona) | 8,402 | 4,429 | 101 | 0 | 1,839 | 0 | 0 | 0 | 2,033 | 0 | 3,474 |
| DUGiFonsEspecials (Universitat de Girona) | 720 | 720 | 0 | 0 | 0 | 0 | 0 | 0 | 0 | 0 | 65 |
| DUGiMedia (Universitat de Girona) | 2,789 | 0 | 0 | 0 | 0 | 0 | 0 | 0 | 0 | 2,789 | 7 |
| Edinburgh Research Archive (University of Edinburgh) | 5,524 | 212 | 257 | 46 | 4,930 | 1 | 0 | 0 | 68 | 0 | 0 |
| ELEA (Universita Degli Studi di Salerno) | 1,043 | 133 | 0 | 1 | 472 | 0 | 0 | 0 | 513 | 0 | 967 |
| espace @ Curtin (Curtin University of Technology) | 16,610 | 6,131 | 1,089 | 4 | 1,952 | 0 | 0 | 0 | 3,960 | 0 | 0 |
| ETDs Repository (VŠKP - University of Economics, Prague) | 37,923 | 166 | 0 | 1 | 37,757 | 0 | 0 | 0 | 0 | 0 | 11 |
| Fraunhofer ePrints (Fraunhofer Gesellschaft) | 12,884 | 1,742 | 578 | 4,734 | 744 | 0 | 0 | 0 | 4,743 | 1 | 0 |
| Ghent University Academic Bibliography (Ghent University) | 22,623 | 10,409 | 950 | 7,638 | 2,034 | 0 | 0 | 81 | 2,048 | 0 | 25 |
| Infoscience (École Polytechnique Fédérale de Lausanne (EPFL)) | 30,762 | 30,762 | 0 | 0 | 0 | 0 | 0 | 0 | 0 | 0 | 4,164 |
| Iowa Research Online (University of Iowa Libraries) | 16,303 | 1,590 | 188 | 65 | 2,393 | 34 | 0 | 57 | 12,037 | 256 | 24 |
| Kent Academic Repository (University of Kent) | 4,270 | 2,045 | 387 | 0 | 75 | 6 | 0 | 38 | 1,695 | 0 | 0 |
| Leeds Met Open Search (Leeds Metropolitan University) | 3,546 | 1,694 | 611 | 122 | 12 | 15 | 0 | 13 | 1,113 | 0 | 342 |
| Leiden University Repository | 14,675 | 8,638 | 1,677 | 76 | 2,574 | 0 | 0 | 157 | 661 | 0 | 8 |
| Lirias (KU Leuven Association) | 26,011 | 12,736 | 1,753 | 5,467 | 1,309 | 7 | 0 | 410 | 4,323 | 1 | 866 |
| Loughborough University Institutional Repository | 8,008 | 4,866 | 157 | 824 | 592 | 3 | 0 | 23 | 786 | 25 | 0 |

| | | | | | | | | | | |
|---|---|---|---|---|---|---|---|---|---|---|
| LSE Learning Resources Online (London School of Economics and Political Science) | 152 | 152 | 0 | 0 | 0 | 0 | 0 | 0 | 0 | 0 | 170 |
| LSE Research Online (London School of Economics and Political Science) | 12,044 | 2,615 | 4,880 | 374 | 8 | 0 | 0 | 0 | 4 | 0 | 7 |
| LSE Theses Online (London School of Economics and Political Science) | 969 | 0 | 0 | 0 | 969 | 0 | 0 | 0 | 0 | 0 | 1,467 |
| LSHTM Research Online (London School of Hygiene and Tropical Medicine) | 6,208 | 5,636 | 8 | 44 | 368 | 0 | 0 | 69 | 323 | 11 | 0 |
| MADOC Publikationsserver (Mannheim University Library) | 4,037 | 137 | 32 | 86 | 551 | 0 | 0 | 115 | 3,118 | 0 | 29 |
| Manchester eScholar (Manchester University) | 5,891 | 3,764 | 114 | 948 | 735 | 0 | 0 | 0 | 318 | 1 | 0 |
| National Chung Hsing University Institutional Repository (National Chung Hsing University, Taiwan) | 67,245 | 17,499 | 117 | 729 | 28,168 | 0 | 0 | 0 | 20,733 | 0 | 0 |
| Open Access LMU (Universitätsbibliothek der LMU München) | 18,547 | 9,152 | 2,258 | 1,073 | 295 | 0 | 0 | 0 | 5,769 | 0 | 169 |
| Opus: Online Publications Store (University of Bath) | 4,834 | 2,711 | 304 | 430 | 945 | 0 | 0 | 0 | 419 | 0 | 1,408 |
| ORBi (Open Repository and Bibliography) (University of Liège) | 39,981 | 19,665 | 2,470 | 10,449 | 1,084 | 0 | 47 | 592 | 4,474 | 0 | 0 |
| OUR@oakland (Oakland University) | 3,185 | 0 | 28 | 0 | 36 | 4 | 0 | 0 | 162 | 48 | 0 |
| Portal de Revistas PUCP (Pontificia Universidad Catolica del Peru) | 10,179 | 10,179 | 0 | 0 | 0 | 0 | 0 | 0 | 0 | 0 | 103 |
| PUCRS Institutional Repository (Pontifical Catholic University of Rio Grande do Sul) | 5,759 | 1 | 0 | 0 | 1 | 0 | 0 | 0 | 5,757 | 0 | 4 |
| Repositori Digital de la UPF (Universitat Pompeu Fabra) | 4,208 | 373 | 18 | 79 | 280 | 0 | 0 | 4 | 2,081 | 0 | 176,032 |
| Repositorio Digital de Tesis PUCP (Pontificia Universidad Catolica del Peru) | 2,847 | 0 | 0 | 0 | 2,847 | 0 | 0 | 0 | 0 | 0 | 0 |
| Repository Utrecht University | 66,879 | 38,717 | 8,165 | 972 | 5,299 | 0 | 725 | 6,469 | 5,638 | 0 | 0 |
| Research Repository (RMIT University) | 3,124 | 634 | 46 | 438 | 1,996 | 0 | 0 | 0 | 10 | 0 | 0 |
| RiuNet (Universitat Politècnica de València) | 27,036 | 5,215 | 211 | 534 | 6,500 | 0 | 5 | 29 | 14,623 | 0 | 2 |
| ROAR (University of East London Repository) | 2,383 | 910 | 200 | 464 | 681 | 0 | 0 | 4 | 141 | 5 | 0 |
| Scholar Commons (University of South Florida) | 17,264 | 0 | 0 | 0 | 0 | 0 | 0 | 0 | 17,264 | 0 | 0 |

| | | | | | | | | | | |
|---|---|---|---|---|---|---|---|---|---|---|
| ScholarWorks @ UVM (University of Vermont) | 863 | 0 | 0 | 0 | 0 | 0 | 0 | 0 | 863 | 0 | 0 |
| SDEIR (Université du Québec à Chicoutimi) | 600 | 0 | 0 | 0 | 0 | 0 | 12 | 0 | 212 | 0 | 81 |
| SHAREOK Repository (University of Oklahoma/Oklahoma State University) | 13,776 | 45 | 17 | 0 | 6,493 | 0 | 0 | 0 | 7,221 | 0 | 0 |
| Swinburne ImageBank (Swinburne University of Technology) | 3,322 | 0 | 0 | 0 | 0 | 3,281 | 0 | 0 | 29 | 12 | 0 |
| Sydney eScholarship Repository (University of Sydney) | 5,118 | 0 | 40 | 0 | 3,371 | 4 | 0 | 0 | 1,607 | 22 | 0 |
| TEORA (Telemark University College) | 1,093 | 282 | 9 | 0 | 237 | 2 | 0 | 0 | 559 | 0 | 0 |
| The Portal to Texas History (University of North Texas) | 281,388 | 0 | 0 | 0 | 0 | 85,342 | 15,276 | 0 | 180,678 | 92 | 74 |
| ThinkTech (Texas Tech University) | 19,430 | 570 | 2 | 1 | 17,622 | 164 | 8 | 0 | 880 | 15 | 0 |
| UBIRA ePapers (University of Birmingham) | 1,138 | 35 | 300 | 0 | 0 | 463 | 0 | 0 | 258 | 1 | 0 |
| UBIRA eTheses (University of Birmingham) | 4,397 | 0 | 0 | 0 | 4,397 | 0 | 0 | 0 | 0 | 0 | 0 |
| UCL Discovery (University College London) | 19,899 | 12,484 | 495 | 1,444 | 3,913 | 0 | 0 | 1 | 1,524 | 0 | 2,907 |
| Unitec Research Bank (Unitec Institute of Technology) | 1,344 | 330 | 0 | 178 | 607 | 0 | 0 | 1 | 128 | 0 | 0 |
| University of Guelph Theses and Dissertations | 1,928 | 0 | 0 | 0 | 1,928 | 0 | 0 | 0 | 0 | 0 | 0 |
| UNT Digital Library (University of North Texas) | 57,511 | 556 | 1,115 | 2 | 5,455 | 7,342 | 26,451 | 18 | 14,883 | 286 | 0 |
| VCU Scholars Compass (Virginia Commonwealth University) | 758 | 623 | 15 | 23 | 0 | 0 | 0 | 1 | 96 | 0 | 0 |
| VU-DARE (VU University Amsterdam) | 24,373 | 16,271 | 786 | 1 | 3,006 | 0 | 0 | 227 | 3,827 | 0 | 376 |
| White Rose Research Online (White Rose University Consortium) | 8,091 | 5,793 | 1,568 | 661 | 0 | 0 | 0 | 0 | 12 | 0 | 0 |
| ZORA (University of Zurich) | 27,146 | 19,152 | 247 | 1,706 | 1,492 | 0 | 0 | 4 | 3,790 | 0 | 0 |
| **TOTAL** | **1,601,063** | **348,102** | **41,370** | **49,487** | **381,222** | **102,358** | **42,533** | **10,160** | **400,316** | **6,638** | **209,440** |